# A Method for Finding Communities of Related Genes

Dennis Wilkinson and Bernardo A. Huberman*
Stanford University and HP Laboratories
1501 Page Mill Rd., Palo Alto, CA, 94394

## Abstract

We present an automated method of identifying communities of functionally related genes from the biomedical literature. These communities encapsulate human gene and protein interactions and identify groups of genes that are complementary in their function. We use graphs to represent the network of gene cooccurrences in articles mentioning particular keywords, and find that these graphs consist of one giant connected component and many small ones. In addition, the vertex degree distribution of the graphs follows a power law, whose exponent we determine. We then use an algorithm based on betweenness centrality to identify community structures within the giant component. The different structures are then aggregated into a final list of communities, whose members are weighted according to how strongly they belong to them. Our method is efficient enough to be applicable to the entire Medline database, and yet the information it extracts is significantly detailed, applicable to a particular problem, and interesting in and of itself. We illustrate the method in the case of colon cancer and demonstrate important features of the resulting communities.

# 1. Introduction

It is well established that the amount of biological information is growing rapidly, and that bioinformatics plays an important role in organizing and analyzing it. An area where the techniques of bioinformatics are particularly applicable is the problem of understanding the function of human genes and proteins. This subject is complex because of the large number of human genes - at least 30,000 - and the interrelated nature of their function. The online biomedical database Medline[1] already contains abstracts for over 12 million articles, more and more of which discuss gene and protein function. It is simply not feasible for researchers to keep abreast of the latest developments by reading articles. Automated information extraction from the literature is needed to organize the available information, make it easily accessible, and to provide insight into directions of future research.

Natural language processing techniques have been used to extract detailed information on gene and protein interactions from text [1-7]. However, they are too computationally intensive to be applied on a large scale, and can only focus on a small subset of genes and/or articles. A different approach is to identify only simple information such as gene and protein names in each article [8-10]. The recent appearance of online libraries of gene and protein names has simplified this process, permitting rapid term extraction from articles and analysis of this data. Stephens et al [11] used the "term-frequency, inverse document frequency" metric to find relations between certain gene and protein names extracted from several thousand Medline documents. Stapley and Benoit [12] extracted yeast gene names from 2500 Medline documents containing the words '*Saccharomyces cerevisiae*' and used a simple metric based on cooccurences to identify pairs of closely related genes. On a larger scale, Jenssen et al [13] extracted human gene names and symbols from all Medline titles and abstracts to create a network of gene cooccurrences, and showed that co-occurring genes were biologically related. Most recently, Adamic et al [14] introduced a fast, accurate technique for identifying genes statistically related to particular diseases. They mined all of Medline for gene name symbols, and used a disambiguation process to reduce the drastic overcounting errors caused by false positives.

Useful as these methods are, they still leave unresolved a finer grained level of detail, i.e. which genes need to be coexpressed. This complementarity is important because the number of genes involved in many cellular processes is very large. The information required to express the complementarity of genes is implicit in the literature; all that is required is a suitable method to extract it and present it in a coherent way. This is the problem we address in this paper.

In particular, we present a method which identifies communities of related genes from the Medline database. These communities are designed to be faithful to known gene and protein interactions, but they are not meant to perfectly model every cellular process. Instead, they are created in order to identify genes which the literature suggests are (or are not) likely to be related, to bring to light possible relations between lesser-known genes, and to identify possible connections between large groups of interacting genes. They are tools which the biologist can use to summarize available information and rapidly narrow his search for new and interesting interactions. They also reveal interesting properties of the literature network which warrant investigation.

Our method allows us to place "ambiguous" genes in more than one community,

---

[1] The search utility for Medline is PubMed at http://www.ncbi.nlm.nih.gov/entrez/query.fcgi

which gives a quantitative estimate of how strongly each gene belongs to each community. Some genes may belong in more than one community, based on their participation in different cellular processes. Also, some communities which our method separates may be related to one another, and "shared" genes would indicate this.

The process is fast and flexible, and can be easily modified to focus on different genes by varying the search keywords. The results are easy to understand and immediately applicable to current biological research.

Method Overview

Following [14], we first extract gene name symbols from each article title and abstract. We identify genes which are statistically relevant to a particular set of keywords, and create a network of cooccurrences of these genes. We represent the network by a graph, which turns out to consist of one giant connected component and many small components. The next task to discover distinct communities of related genes within the giant component. This is achieved by exploiting notions of betweenness centrality, which allow us to efficiently partition the graph. We repeatedly partition the graph to create different community structures, and aggregate them into one final list of communities.

Section 2 of this paper describes the process of extracting gene mentions and determining which genes are relevant to a set of keywords. Section 3 discusses the construction of a gene cooccurrence graph and its properties. In Section 4 we explain the method for the partitioning of the giant component. Results using keywords related to colon cancer are presented in Section 5, and Section 6 summarizes the strengths of the method and discusses possible improvements.

## 2. Obtaining Cooccurrence Data

As stated above, we identify literature cooccurrences of genes relevant to a disease, using the method of [14].

Using a list of all official and alias symbols for human genes compiled from the HUGO[2], OMIM[3], and Locuslink[4] web sites, we automatically extracted the gene name symbols and disease mentions from all Medline article titles and abstracts. Where possible, we replaced alias symbols with official ones. We also extracted keywords related to a certain disease, and used them to determine which genes were statistically correlated with this disease. Restricting the network to these genes ensures that it is small enough to work with but large enough and dense enough to contain interesting information. Keywords related to other cellular processes could also be used.

To test a gene for statistical relevance to a disease, we simply compared the observed number of gene-disease cooccurrences to the number we would expect given no correlation. Since the distribution of cooccurrences of two uncorrelated terms follows a binomial distribution, a value of observed gene-disease cooccurrences more than one standard deviation greater than the binomial expected value indicates correlation. This statistical method is preferable to the "term frequency, inverse document frequency" metric because it accurately handles infrequently mentioned genes, which are very common.

The final step in obtaining data was to remove false positives, which occur frequently because gene symbols generally coincide with other abbreviations having nothing to do with genes. For example, the symbol HDC, representing the gene "histidine decarboxylase," was commonly used in the literature as an abbreviation for "high dose chemotherapy." We disambiguated the data, using a method shown in [15] which yielded unambiguous symbol identifications with very low error rate.

---

[2] http://www.gene.ucl.ac.uk/nomenclature/
[3] http://www3.ncbi.nlm.nih.gov/omim/

## 3. Gene Graph

The creation of gene graphs from the cooccurrence data was performed following a well-known procedure [12,13]. Each vertex in the graph represents a gene, and an edge exists between two vertices if the genes they represent cooccur at least once. We did not use weighted edges. In creating the graph, we neglected articles published before 1990 and articles which listed more than 5 genes.[5]

The resulting graph has a power law distribution in its degree. That is, the number of vertices of degree x is given by $Ax^{-\beta}$, where $\beta > 0$. This is shown in Figure 1, where we plot the data on a log-log scale for gene graphs corresponding to several diseases.

The properties of such power law graphs have been extensively studied [16,17]. We expect the graph to consist of one giant connected component, and other small components of size $O(\ln(N))$, since $2 < \beta < 3.5$ [17]. Here $N$ is the size of the graph and $\beta$ is the power law exponent. The sizes of the connected components in graphs corresponding to several diseases, in agreement with [17], are shown in Table 1.

---

[4] http://www.ncbi.nlm.nih.gov/LocusLink/

[5] Older articles are frequently a source of error because of nomenclature differences. In addition, they contribute little because fewer genes were known then, and the important ones are discussed in more recent articles anyway. Articles in which many genes cooccurred were often "survey" type articles which did not discuss gene interactions. See section 4.

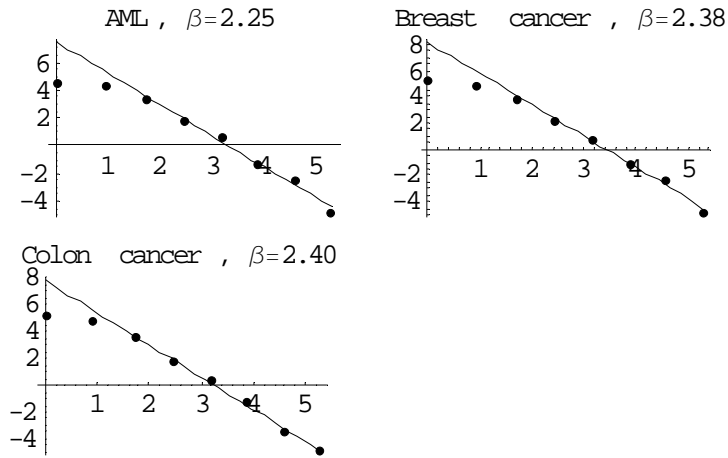

Figure 1. The number of vertices (y-axis) is plotted against the degree of the vertex (x-axis) for several diseases on a log-log scale. We followed the usual binning procedure in plotting the data. The deviation from the power law for low vertex degree is typical.

| **AML** (total 488 genes) | |
|---|---|
| size of component | number of such components |
| 460 | 1 |
| 4 | 1 |
| 3 | 4 |
| 2 | 6 |

| **Breast cancer** (816 genes) | |
|---|---|
| size of component | number of such components |
| 686 | 1 |
| 6 | 2 |
| 5 | 1 |
| 4 | 5 |
| 3 | 9 |
| 2 | 33 |

| **Colon cancer** (682 genes) | |
|---|---|
| size of component | number of such components |
| 561 | 1 |
| 4 | 4 |
| 3 | 15 |
| 2 | 30 |

Table 1. Sizes of connected component in several gene graphs.

Since the smaller components contain few genes with few neighbors, they are of limited interest. They usually consist of little-known genes which have not been related to other genes. In what follows, we focus exclusively on the giant component.

## 4. Community Structure

There is no formal definition for a community of vertices within a graph. A graph can be said to have community structure if it consists of subsets of genes, with many edges connecting vertices of the same subset, but few edges lying between subsets [18]. Finding communities within a graph is an efficient way to identify groups of related vertices.

In order to explain the community discovery process, which is based on that of Girvan and Newman [18], we consider as a first example the small graph shown in Figure 2. This graph consists of two well-defined communities: the four vertices denoted by squares, including vertex A, and the nine denoted by circles, including vertex B.

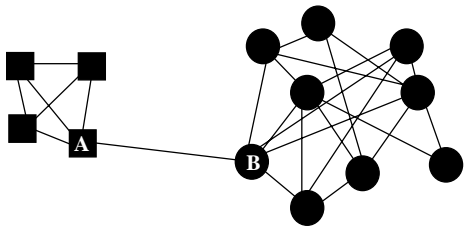

Figure 2.

In the context of Figure 2, edge AB has the highest betweenness. If we were to remove it, the graph would split into two connected components, the square and circle communities. This illustrates the idea behind our method of imposing community structure on a graph: we repeatedly identify inter-community edges of large betweenness and remove them, until the giant component is resolved into many separate communities.

To find inter-community edges, we exploit Freeman's [19] idea of betweenness centrality, or betweenness, applied to edges. The betweenness of an edge is defined as the number of shortest paths that traverse it. This property distinguishes inter-community edges, which link many vertices in different communities and have high betweenness, from intra-community edges, whose betweenness is low.

The removal of an edge strongly affects the betweenness of many others, and so we must repeatedly recalculate the betweenness of all edges. To do this quickly, we used the fast algorithm of Brandes [20], whose basic strategy is the following. Consider the shortest paths between a single vertex, the "center", and all other vertices. Calculate the betweenness of each edge due only to these shortest paths, and add them to a running total. Then change centers and repeat until every vertex has been the center once. The running total for each edge is then equal to exactly twice the exact betweenness of that edge, because we have considered all the pairs of endpoints of paths twice.

Our procedure stops removing edges when we cannot further meaningfully subdivide our communities; for example, as in Figure 2, after removing edge AB. What criterion tells us when to stop? As we remove edges, we divide the graph into many unconnected components. Structurally, a component of 5 or fewer vertices cannot consist of two viable communities. The smallest possible such component is size 6, consisting of two triangles linked by one edge (Figure 3). If at any time we remove an edge from our graph and separate a component of size < 6, we can identify it as a community.

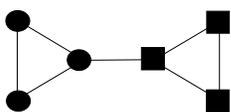
Figure 3.

Components of size ≥ 6 can also be individual communities, like the group of 9 in Figure 2. To identify this type of component as a community, we use an intuitive threshold based on the betweenness of an edge connecting a <u>leaf</u> vertex, or vertex of degree one, to the rest of the graph. Consider the graph of Figure 4 below. It is clear that it consists of just one community. Applying the Brandes algorithm, we find that edge XY has the highest betweenness, indicating that the size of the largest distinct community within the graph has size 1. That is, there are no distinct communities within the graph.

In general, the single edge connecting a leaf vertex (such as X in Figure 4) to the rest of a graph of $N$ vertices has a betweenness of $N-1$, because it contains the shortest path from X to all $N-1$ other vertices. The stopping criterion for components of size $\geq 6$ is therefore that the highest betweenness of any edge in the component be equal to or less than $N-1$.[6]

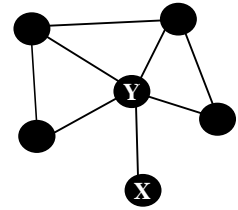

Figure 4.

We can now explain the need to neglect survey-type articles which list many genes in creating our graph. The genes listed in these articles will all be linked to one another, forming what is known as a complete subgraph $K_n$. Such a grouping is very tightly knit and will likely not be split into different communities. This situation could be due only to the survey article, not accurately reflecting the interactions between the genes. It is possible that a few articles mention many genes which are in fact functionally related, but we have good reason to hope that in this case the genes are discussed in smaller groups in other articles.

Communities Consist of Functionally Related Genes

The communities thus created consist of genes that were strongly interrelated in the literature. Most, but not all, gene cooccurrences imply a functional relation; genes may also cooccur in an article abstract because of physical proximity, similarity of nomenclature or structure, historical association, or other reasons. However, since such non-functional edges are a minority, they are highly likely to be inter-community, because the neighbors of two non-functionally related genes are unlikely to be linked.

---

[6] It is not in general true that an inter-community edge must have betweenness greater than N-1. For a community of size m within a graph of size N, there is a total betweenness of $m(N-m)$ divided among the edges connecting the community to the graph. So, if there are more than m such edges, it is possible that none of them will have betweenness greater than N. However, remember that none of these edges, or the extra-community vertices they connect, should be adjacent, because then m would not be a community. This type of situation is extremely unlikely in a power law graph. Even in Girvan and Newman's highly non-power law college football graph, the highest betweenness at any step is only occasionally less than $N-1$.

For example, genes S100A4 and S100A6 are members of the S100 family and cooccur twice in articles related to colon cancer, but they are not functionally related.[7] In our results, S100A4 and S100A6 do not occur in a community together. The neighbors of one are not linked to the neighbors of the other, which causes them to be in separate communities.

Multiple Community Structures

The examples given thus far in Figs. 2-4 are small graphs with clear-cut community structure, not at all similar to large, dense gene graphs. Many "ambiguous" genes may be logically placed in two or more communities, in the context of both our graph and of biological function, so one cannot simply impose one rigid community structure on our graph. In the course of determining a community structure on the graph for colon cancer, we removed thousands of inter-community edges. As we will clarify below, the order of removal of edges strongly affects which other edges are removed later.

Consider the subgraph of Figure 5. It consists of two communities, one on the left including vertex A, and another on the right including C. Among its edges, BC initially has the highest betweenness, and AB's betweenness is also high. Once we remove BC, however, AB becomes an intra-community edge with low betweenness, and it will never be removed. Gene B will eventually be placed in a community with gene A. Had we removed AB first, BC would be rendered intra-community, and gene B would end up in the community with C.

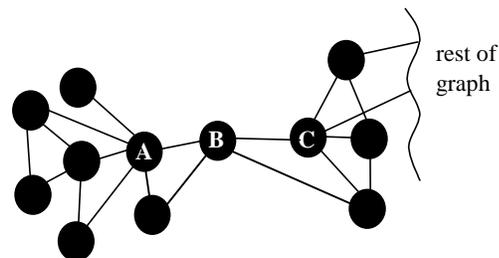

Figure 5.

---

[7] Medline PMIDs 10389988 and 10952782

Moreover, in considering Figure 5, it is not clear where B should end up. In fact, from the graph alone, B is ambiguous, and could rightfully be considered to be a part of both communities. Such ambiguous genes are common in our gene graph.

To account for ambiguous genes, we modified the Brandes algorithm described above to repeatedly obtain reasonable, yet slightly different, community structures. Instead of using every vertex as the "center" once, we cycle randomly through at least m centers (where m is some arbitrary cut-off), until the betweenness of at least one edge exceeds a threshold, again based on the betweenness of a "leaf" vertex. We then remove the edge whose betweenness is highest at that point, and repeat until we have broken the graph into communities, identified by the criterion described above. We only do this for large components, using the full Brandes algorithm for small ones. See Appendix 1 for more detail. The modified algorithm may occasionally remove an intra-community edge, but such errors are unimportant when one compares a large number of community structures.

Applying this modified process n times, we obtain n community structures imposed on the graph. We can then compare the different structures and identify communities, as well as the strength of each gene within the community. For example, after imposing 45 structures on our graph, we might find: a community of genes A, B, C, and D in 20 of the 45 structures; a community of genes A, B, C, D, and E in another 20; and one of genes A, B, C, D, E and F in the remaining 5. We report this result in the following way

A(45) B(45) C(45) D(45) E(25) F(5)

which signifies that A, B, C, and D form a well-defined community, E is related to this community, but also to some other(s), and F is only slightly, possibly erroneously, related to it.

To aggregate our different community structures in this way, we used a modified intersection/union metric which identifies a closest match for each community of two structures. We chose to compare each structure individually to a "master list" which was updated as we progressed. As an illustration, returning to the above example, all communities imposed our 46th structure would be compared to {A(45), B(45), C(45), D(45), E(25), F(5)}. The intersection/union metric was weighted according to how strong a gene's presence in that community was. See Appendix 2 for a more detailed description. The entire process of determining community structure is displayed in Table 2.

```
A. For i iterations, repeat {
        1. Break the graph into connected components.
        2. For each component, check to see whether component is a
    community.
                a. If so, remove it from the graph and output it.
                b. If not, remove edges of highest betweenness, using the
                   modified Brandes     algorithm for large components, and the
                   normal algorithm for small ones. Continue removing edges
                   until the community splits in two.
        3. Repeat step 2 until all vertices have been removed from the graph
           in communities.
}
B. Aggregate the i structures into a final list of communities.
```
Table 2. Algorithm for determining community structure

# 6. Results

We applied the above technique using keywords related to colon cancer. We considered articles which mentioned at least one of "colon," "colorectal," colonic," or "gastrointestinal," and at least one of "cancer" or "carcinoma." We identified 682 genes which were statistically correlated with colon cancer and which cooccured in these articles with at least one other correlated gene. The graph of this cooccurrence network consisted of a giant component of 561 genes, and other uninteresting smaller components (Table 1). The community discovery algorithm split the giant component into 79 different communities, with sizes ranging from 2 to 50 genes. 29 of the communities were "small", consisting of 4 genes or less; 7 were "large", containing 25 or more genes; and the other 44 were "medium-sized."

We quickly found that medium-sized communities were the most useful, because they were easy to analyze and contained interesting information. Forty-four of them from this search give plenty of results to analyze. As we mentioned in the introduction, it is not the purpose of this method to produce communities which perfectly model certain cellular processes, or to exactly reproduce known groups of genes whose interactions are well-understood. Rather, our method allows researchers to quickly draw conclusions about genes being studied in the context of a given disease.

Our results, along with Medline and OMIM, allow us to suggest undocumented connections between genes of one community, and between genes in different communities. They also demonstrate that our method tends to separate genes which cooccurred, but were not functionally related, into different communities, as discussed in

Section 3.

In this section we discuss one community in detail, and present results from several others. A full list of colon cancer communities will be published on our website, in the format we show below. We will additionally include a utility which allows the user to find the article abstracts in which any two genes cooccurred, referenced by Medline PMID number.

| gene symbol | score in community | overall mentions with colon cancer | neighbors with colon cancer (star denotes neighbor not in community) |
|---|---|---|---|
| PTGS2 | 50 | 263 | PTGS1* DLD* MLH1* BCL2* PLA2G2A PLA2G4 APC* ERBB2* PGES ERBB3* PLA2 ACL4 WNT1* GRP* GRPR* LEF DLR* TCF4* TCF* MYB* VEGF* NOS2A TP53* MADH4* EGFR* S11* PDCD4 BRCA1* BRCA2* MSH2* ERBB4 |
| PLA2G2A | 50 | 12 | APC* PTGS2 PLA2G4 TP53* NF2* DCC* MLH1* SPLA2 |
| PLA2G4 | 50 | 1 | PLA2G2A PTGS2 |
| SPLA2 | 50 | 4 | PTGS2 PLA2G2A |
| FACL4 | 50 | 1 | PTGS2 |
| NOS2A | 50 | 7 | PTGS2 |
| PDCD4 | 50 | 1 | PTGS2 |
| PGES | 18 | 2 | ERBB2* PTGS2 ERBB3* |
| LEF1 | 5 | 13 | WNT1* TCF* PTGS2 TCF4* APC* FRA1* PLAUR* MYC* MMP7* TCF7* |

Table 3. A sample community of 9 genes from our results for colon cancer. Here "score in community" denotes how many times the gene was placed in this community (see section 4 on multiple community structures).

Table 3 shows a community of genes related to colon cancer. Genes in this community are related to the overexpression of PTGS2, prostaglandin-endoperoxide synthase 2, in colon cancer. Although PTGS2 is the official HUGO symbol, this gene is very commonly called COX-2 (cyclooxygenase-2), and we will use this term.

The features of this community suggests the following possibilities: connections between some of the genes which cooccur with COX-2, but not each other; good reasons why many of the neighbors of COX-2 are not in this community; and possible connections to other communities via PGES and LEF1. We investigated these possibilities and present the results below.

Unexplored connections

We immediately found a possible unexplored connection between the phospholipase A2 genes in this group (SPLA2, PLA2G4 (aka cPLA(2)), PLA2GA2) and FACL4, via COX-2 and arachidonic acid. COX enzymes convert arachidonic acid to prostaglandins[8]. The three phospholipase A2 genes in the group are all sources of arachidonic acid [9, 10], and are thus related to COX-2. However, we found that the FACL4 enzyme also uses arachidonic acid, and that "the cellular level of unesterified arachidonic acid is a general mechanism by which apoptosis is regulated and that COX-2 and FACL4 promote carcinogenesis by lowering this level."[11] This indicates a connection between FACL4 and the phospholipase A2 family of genes in carcinogenesis. However, a Medline search for FACL4 or its alias ACS4 with each of PLA2, SPLA2, PLA2G4 and PLA2G2A turned up no results. The OMIM entry for FACL4 has no mention of phospholipase A2. It would have been very difficult for a researcher to discover this connection manually from Medline; even a search for "arachidonic acid" and "colon cancer" together produces 119 abstracts to sift through. Additionally, during this brief literature search we discovered that nonsteroidal antiinflammatory drugs (NSAIDs) function by suppressing cPLA2 (PLA2G4) mRNA expression and thus depriving COX-2 of arachidonic acid.[10] Our method therefore suggests that these drugs also affect FACL4 expression, although a Medline search of NSAID and FACL4 turns up no results.

---

[8] 11274413, for example
[9] Medline PMID 11274413, for example
[10] 11789254, for example
[11] 11005842

Absent neighbors

In examining neighbors of PTGS2 (COX-2) not present in this community, we noticed in particular the similarly named gene PTGS1 (aka COX-1). These two genes are isoforms of cyclooxygenases[12], and cooccured 70 times in articles in connection with colon cancer. In fact, they have been shown to regulate colon carcinoma-induced angiogenesis by two different mechanisms[13]. COX-2 has also been shown to be expressed much more frequently than COX-1 in tumors, and less frequently in normal tissue[14]. There is thus a good reason why these two genes always ended up in different communities. Although the enzymes they code for are structurally very similar, COX-2 plays a strong role in colon cancer, while COX-1's role is weaker and by a different mechanism. It is unlikely that a neighbor of COX-2 is related to COX-1, although they both cooccur with COX-2.

Several other neighbors of PTGS2, such as MLH1, BRCA1, BRCA2, and MSH2, also proved to be weakly or non-functionally related. However, a few of PTGS2's non-community neighbors have been tentatively identified as functionally related, such as GRP and GRPR (GRP receptor) [15], and EGFR [16]. For this reason we include a list of all neighbors of each gene in the results, as a secondary list of possible connections to explore.

---

[12] 9099957, for example
[13] 9630216
[14] 7780968, for example. Note the use of the alias PGHS-1 or -2 for COX-1 or -2 in this article.
[15] 11292836

Links to other communities

We also looked for links to other communities through the genes PGES and LEF1, both of which are often placed in other communities.

Both genes yielded good results. PGES only cooccurs with other genes once, in an abstract with COX-2, ERBB2 and ERBB3 . Examining this abstract, we find a link between the COX-2 pathway and autocrine/panacrine activation of HER2/HER3 (aka ERBB2 and ERBB3) [17]. The ERBB genes are present in another community of 25 genes. This article links all the genes related to arachidonic acid, most of which do not cooccur with ERBB2 or 3, to genes of the ERBB2/ERBB3 community.

LEF1 was found with COX-2 in only one article[18]. It states that, "NO [nitric oxide] may be involved in PGHS-2 [COX-2] overexpression in conditionally immortalized mouse colonic epithelial cells. Although the molecular mechanism of the link is still under investigation, this effect of NO appears directly or indirectly to be a result of the increase in free soluble beta-catenin and the formation of nuclear beta-catenin/LEF-1 DNA complex." This article definitely indicates a connection between COX-2, NOS2A (nitric oxide synthase, responsible for the production of NO), and the very important colon cancer gene beta catenin.

As a last note, this community demonstrates the crucial importance of considering alias symbols when extracting gene names. The aliases COX-2, PGHS-2, NOX2, and

---

[16] 9012840
[17] 9927187

cPLA(2) all played essential roles in tying it community together.

Other Notable Results

Here we present several other notable results we found after a morning of searching through our communities.

One medium-sized community indicates a connection between PXR (pregnane X receptor) and GP170 (P-glycoprotein). PXR is implicated in the induction of the MDR1 gene[19], while MDR1 expression has been associated with the expression of functional P-glycoprotein.[20] A Medline search turns up no results for gp170 or gp-170 with pxr or its aliases par, sxr, and nr1i2.

The small community 27, although it consists of only 4 genes, turns up a probable, undocumented connection between gp200-MR6 and STAT6, via IL-4 and its receptor IL-4R. IL-4 induces STAT6, which is involved in mediating activation of IL-4R gene expression,[21] while gp200-MR6, has been shown to be functionally associated with IL-4R.[22] This example is notable because the articles mentioning colon cancer where the genes cooccur, which create the community, are relatively old: PMIDs 8530527, from *J Biol Chem*, Dec. 22, 1995, and 9178815, from *Int J Cancer*, May 16, 1997 respectively.

While large communities are more difficult to analyze for the non-expert, we were nevertheless able to draw some conclusions. For example, we considered a 30-gene community largely concerned with apoptosis and genes related to BCL-2, containing in particular the gene TRAIL. TRAIL has been shown to induce procaspase-8 activation, triggering caspase-dependent apoptosis in colon cancer cells.[23] It could thus be related to the function of genes such as BCLX, BCLXS, etc, which we find in this community but which do not cooccur with TRAIL, via the genes BCL-2 and CASP8.

---

[18] 10834941
[19] 11297522
[20] 10334913
[21] 8810328

Another good examples of neighboring genes with similar names placed in different communities for good reason is MMP11 and MMP9[24]. Often, non-functionally related neighboring genes have the majority of their counts in different communities; examples of this include CYP3A4 or CYP3A5 and CYP1A2,[25] as well as SMAD3 and SMAD5.[26]

## 7. Conclusions

We have presented a data mining technique for biological literature which produces detailed results while extracting only very simple data from each article abstract and title. The method produces a list of communities of functionally related genes which are designed to summarize available information and indicate genes which are likely to be complementary in their function. The genes within a community are weighted, indicating how strongly they belong to the community. We show that the communities produced in the case of colon cancer have interesting features which give one insight into the function of the component genes.

The existence many community structures on each gene graph implies that genes are a part of several different communities. We thereby produce a much richer result than if we had imposed one rigid structure on the graph. This idea could be applied to social and other networks where individuals play a role in more than one community.

---

[22] 9178815
[23] 11245478
[24] 8645587
[25] 9202751
[26] 10446110 and 11196171, for example. SMAD2 and SMAD4 are aliases for MADH2 and MADH4, respectively.

We introduce two statistical components into the process, which lessen the inevitable errors of text mining in the biological literature, which are particularly severe in our case because of the complex, young nomenclature system for genes. However, our method retains the ability to detect relations between rarely-mentioned genes, one of its strongest features.

The factor which most limits our results is the absence of many gene symbols from HUGO and other online databases. Hopefully, these databases will soon be more complete. Even then, it will be difficult to quickly detect all gene symbols because of small modifications introduced by many authors, such as the problematic addition of hyphens, parentheses or spaces into the symbol.

Another limiting factor was the placement of many genes in either large and very small communities in our results. While still a step forward from raw cooccurrence data, such communities are of limited usefulness. Small communities often did not provide much insight into the function of their component genes, other than that the genes were rarely related to others in the context of colon cancer. If such genes were more commonly mentioned in other contexts, a search using other diseases or keywords would likely turn up more interesting communities with these genes. Large communities were difficult for us to analyze, but nevertheless yielded some interesting results. These communities contained many of the most commonly mentioned genes in connection with colon cancer, such as APC and TP53. Strangely, a search for colon cancer genes is probably not the most efficient way to study these genes, which are simply too highly linked in this context. Instead, one could perform other searches with other keywords, hoping to focus on particular aspects of these genes' function by confining them to smaller, more informative communities.

We believe that large communities are a product of the graph topology, not of the threshold by we use to stop subdividing a community or of the aggregation process. To further subdivide large communities, one could consider a weighted graph, where the weight corresponds to the (normalized) number of times the two genes cooccur. This could increase the "distance" between, for example, two commonly studied, distantly related, cooccurring genes. They would then not end up in the same community, and more importantly not "glue" a false community together. The simplest such weighting would be to neglect all links below some (normalized) threshold weight. Another resolution to the problem of large communities would be to refine the step which aggregates the community structures into one result.

**Acknowledgement**: We thank Lada Adamic, Eytan Adar and Melissa Wilkinson for many useful discussions.

**Appendices**

Appendix 1: Parameters for modified Brandes algorithm

The modified Brandes algorithm cycles through at least m centers and stops when at least one edge's betweenness exceeds a threshold T. We used the normal algorithm, fully calculating betweenness and removing the edge of highest betweenness, for

components of size N<50. For components of N>50:

The cutoff $m(N) = 10\log(N) - 25$. This function has $m(50) \approx 15$, and $m(800) \approx 41$. Fifteen is a reasonable number of centers to consider in for a component of size 50, while 40 centers should be more than enough for any component, however large. An intra-community edge will be erroneously removed if we repeatedly choose centers from the same community. For a component of 50 vertices and 4 communities, the probability of choosing 8 out of 15 centers from one community is about 1%. For a large component with many communities, the probability of error is very low for a cutoff of 40 centers.

The threshold $T = N + i - 2$, where i is the number of centers we have considered. For any i, this is the value of the betweenness of the edge connecting a leaf vertex to the rest of the graph. For low i, we can expect the edge connecting a leaf vertex to have the highest betweenness, once the leaf it connects gets chosen to be a center. This is because the betweenness imparted to a leaf edge when its leaf is the center is N-1, the highest possible value for one edge from one center. In general, for low i, edges will have lower betweenness than such leaf edges. As i increases, inter-community edges gain in betweenness more than intra-community edges, unless we have very bad luck in choosing the centers. In light of the cutoff m, we expect not to have bad luck. We thus expect that intra-community edges will stay below T, while inter-community edges eventually increase past T. In practice, we often had to consider many more centers than m before any edge's betweenness exceeded T.

Appendix 2: Intersection/Union Metric and Aggregating Community Structures

A simple comparison metric for communities $A$ and $B$ of genes $\alpha_1, \alpha_2, ..., \alpha_n$ and $\beta_1, \beta_2, ..., \beta_m$, where some of the $\alpha_i$ may be the same as some of the $\beta_i$, is the intersection/union ratio:

$$\frac{A \bigcap B}{A \bigcup B} = \frac{\sum_{i,j} \delta(\alpha_i, \beta_j)}{n + m - \sum_{i,j} \delta(\alpha_i, \beta_j)},$$

where the sums run over all i and j and $n$ and $m$ are the number of genes in communities $A$ and $B$, respectively. The sum with the Kronecker delta is just a way of writing "how many genes $A$ and $B$ have in common." The intersection/union ratio will be larger for a closer match. Comparing each community in one structure to all the communities in the other, we can find the closest match for each one.

Since we had to compare 50 different structures, we chose to maintain a "master list" of communities, and compare each subsequent structure to it using a weighted intersection/union metric explained below. Each community of the structure was matched to one community in the master list by the metric. The master list started out as simply the first structure, chosen arbitrarily. After comparing each new structure to the master list, we updated the master list. In the updating process, if a community in the structure and one in the master list were linked to each other, we incremented the genes they had in common and appended the new genes. As an example, assume we have a community {A, B, C, D} in the structure which is matched to {B(5) C(5) D(3) E(5)} in the master list. We would update this community in the list to become {A(1) B(6) C(6) D(4) E(5)}; that is, A would be appended, B, C, and D would be incremented, and E would remain unchanged. The numbers following each gene in a community are simply how many times the gene has been associated with that community.

The metric for comparing a new community $A$ to one in the master list $B$ must take the weights of the genes in the master list into account. All genes in the new community are assigned weight 1; each gene $\beta_i$ in the master list community is assigned a fractional weight $\rho_i$ depending on how many times it has occurred in $B$. For example, if we have aggregated 20 structures and the gene DCC has occurred 15 times in community 2, its weight in community 2 would be 0.75. The weighted metric is given by

$$\frac{\sum_{i,j}[\rho_j \cdot \delta(\alpha_i, \beta_j)]}{n + \sum_j \rho_j - \sum_{i,j}[\rho_j \cdot \delta(\alpha_i, \beta_j)]}.$$

Occasionally, two or more communities in the structure were matched to one in the master list, and vice versa. In this case, we assumed that the an intra-community edge had been erroneously removed to divide one community into two or more, either in the structure or the master list (in that case, it would have been in the previous structures creating the master list). We thus melded the "divided" communities into one, altering the master list if need be, and then updated the master list as described above. This step could create a problem if we ended up with huge communities at the end, but we found that in general the largest communities in the final result had only ten or fifteen more genes than the largest communities in each individual structure. This incidentally indicates that our edge removal algorithm had a low error rate.